\documentclass[11pt]{article}

\usepackage[dvips]{graphics}
\usepackage{epsfig}

\makeatletter
\renewcommand{\@cite}[2]{\leavevmode%
\hbox{$^{\mbox{\the\scriptfont0 #1}}$}}
\makeatother

\def\degC{\kern-.2em\r{}\kern-.3em C}

\begin{document}
~\\
~\\
\begin{center}
{\Huge Ubiquity of Log-normal Distributions in \\ Intra-cellular Reaction Dynamics}
\end{center}
~\\
\begin{large}
\begin{center}
Chikara Furusawa\footnotemark[1] \footnotemark[5]{}, 
Takao Suzuki\footnotemark[2]{},
Akiko Kashiwagi\footnotemark[1]{},\\
Tetsuya Yomo\footnotemark[1]\ \footnotemark[2]\ \footnotemark[3]\ 
\footnotemark[4] \footnotemark[5]{},
Kunihiko Kaneko\footnotemark[3] \footnotemark[4] \footnotemark[5]{}
\end{center}
\end{large}
\noindent
\footnotemark[1] Department of Bioinformatics Engineering, Graduate School of 
Information Science and Technology, Osaka University, 2-1 Yamadaoka, Suita, Osaka 565-0871, Japan\\
\footnotemark[2] Department of Biotechnology, Graduate School of Engineering, Osaka University, 2-1 Yamadaoka, Suita, Osaka 565-0871, Japan\\
\footnotemark[3] Graduate School of Frontier Biosciences, Osaka University
1-3 Yamadaoka, Suita, Osaka 565-0871, Japan\\
\footnotemark[4] Department of Pure and Applied Sciences, Univ. of Tokyo
Komaba, Meguro-ku, Tokyo 153-8902, Japan\\
\footnotemark[5] ERATO Complex Systems Biology Project, JST , 3-8-1 Komaba, Meguro-ku, \\
Tokyo 153-8902, Japan \\

\vspace{2cm}
\noindent
{\bf Corresponding Author:} Kunihiko Kaneko\\
\hspace{2mm}Department of Pure and Applied Sciences, Univ. of Tokyo,\\
\hspace{2mm}Komaba, Meguro-ku, Tokyo 153-8902, Japan\\
\hspace{2mm}Tel/FAX: +81-3-5454-6746\\
\hspace{2mm}E-mail: kaneko@complex.c.u-tokyo.ac.jp

\newpage
             
\begin{abstract}
The discovery of two fundamental laws concerning cellular
dynamics with recursive growth is reported. First, the chemical
abundances measured over many cells are found to obey a log-normal
distribution and second, the relationship between the average and
standard deviation of the abundances is found to be linear.
The ubiquity of the laws is explored both theoretically and experimentally.
First by means of a model with a catalytic reaction network,
the laws are shown to appear near the critical state with efficient self-reproduction.
Second by measuring distributions of fluorescent proteins in bacteria cells
the ubiquity of log-normal distribution of protein abundances is confirmed.
Relevance of these findings to cellular function and biological plasticity 
is briefly discussed.
\end{abstract}
~

\section{introduction}

The search for universal statistics with regards to fluctuations in
cellular dynamics is
an important topic in biophysics.
Generally, the molecule numbers of the various chemical species (e.g.,
proteins)
change from cell to cell.
Since many intra-cellular reaction processes are finely tuned to
specific
functions, one would initially expect
the number distributions of the involved chemical species
to be sharp in order to suppress fluctuations.
In reality, however, they are far from sharp, and large fluctuations do
occur.
In order to understand how cells can nevertheless function, it is
essential to
gain insight into the statistics of the chemical abundances.

Indeed, fluctuations in cellular processes have extensively been studied
in stochastic gene expressions and signal transduction these days
\cite{Collins2,Arkin,Siggia,MUeda,Paulsson}.
In particular, significant advances have been made
in the study of their distributions
using fluorescent proteins \cite{Elowitz,Collins}.
In light of these recent advances, it is important to search for general
laws that hold for such distributions.

Previously we found a universal power-law distribution in the average
abundances
of chemicals in cells, by using a simple reaction network model
\cite{Zipf}.
The theoretical conclusions were
confirmed with the help of a large-scale gene expression data \cite{Zipf,Ueda,Kuznetsov}.
The above power law concerns the average over all chemical
species
and forms a first step in the study of universal statistics
in cellular dynamics.
As a next step, it is important to explore
universal characteristics with regards to the distribution of each
chemical over the cells.

Here, we report two basic laws for the
number distributions of chemicals in cells that grow recursively.
The first law is a
log-normal distribution of chemical abundances measured over many cells,
and the second law is a
linear relationship between the average and standard deviation of
chemical abundances.
We give a heuristic argument as to why these laws should hold for a cell
with steady growth,
and demonstrate them numerically using a simple model for a cell with an
internal
reaction network. Lastly, the results of an experimental study
confirming the two laws
are presented.

Indeed, the log-normal distribution is clearly different from the
Gaussian
distribution
normally adopted in the study of statistical fluctuations, and has a
much
larger tail
for greater abundances. Hence the
generality of the laws we report is of considerable significance for all
the
statistical studies
of cellular fluctuations, and is essential to understanding
cellular function, adaptation, and evolution.

\section{Heuristic Argument}

Cells contain huge numbers of chemicals that catalyze each other
and form complex networks.
For a cell to replicate itself recursively, a set of chemicals has to be
synthesized from nutrients supplied from the outside
through biochemical processes driven by the same set of chemicals.
Consequently, it is natural to consider an auto-catalytic process
as the basis of biochemical dynamics within replicating cells.

As a very simple illustration, let us consider
an auto-catalytic process where a molecule (or a set of molecules) $x_m$
is replicated with the aid of other molecules.
Then, the growth of the number $n_m(t)$ of the molecule species $x_m$ is
given by
\begin{math}
dn_m(t)/dt=A n_m(t)
\end{math}
with $A$ describing the rates of the reaction processes that synthesize
the molecule $x_m$.  Clearly, this kind of synthetic reaction process
depends
on the
number of the molecules involved in the catalytic process.
At the same time, however, all chemical reaction processes are
inevitably
accompanied by fluctuations arising from the stochastic collisions of
chemicals.
Thus, even when the reactions that synthesize a specific chemical
to subsequently convert it to other chemicals are balanced in a steady
state,
 fluctuation terms will remain.
Consequently, the above rate $A$ has fluctuations $\eta(t)$ around
its temporal average $\overline{a}$
such that
\begin{math}
d n_m(t)/dt =n_m(t)(\overline{a} + \eta(t))
\end{math},
and hence we obtain
\begin{equation}
d\log n_m(t)/dt =\overline{a} + \eta(t).
\end{equation}

\noindent
In other words, the logarithm of the chemical abundances shows
Brownian motion around its mean, as long as $\eta(t)$ is approximated by
random noise.
Accordingly, one would expect the logarithm of the chemical abundances
(i.e.
molecule numbers) to obey a
normal (Gaussian) distribution (this kind of a distribution is known as
a
log-normal distribution)\cite{KK_PRE}.
In contrast to the Gaussian distribution,
the log-normal distribution has a longer tail representing the higher
frequencies
of greater abundances
when plotted in the original scale without taking the logarithm.

In general, at each step of the auto-catalytic process,
a multiplicative stochastic factor $\eta(t) n_m $ can appear.
Consequently, the log-normal distribution of chemical abundances,
rather than the Gaussian distribution, may be
common for cells that reproduce recursively.

Of course, the above argument is too simplistic to describe the dynamics
of actual
cells. For example,
the fluctuations could be suppressed since the
increase in chemical abundance does not continue for ever
due to cell divisions which might thus alter the form of the
distribution.
Furthermore, in a complex biochemical reaction network,
several reaction processes may work in parallel
for the replication of a chemical.
This leads to the addition of fluctuation terms, and
the central limit theorem of probability theory
might imply that the distribution becomes Gaussian.

Hence it is a priori far from clear whether the simple argument to
support the
log-normal distribution is valid or not.  Nevertheless, since the
log-normal distribution is
rather different from the standard Gaussian distribution,
its universality is of great importance for
understanding the fluctuations in cells and
stochastic gene expressions. Here we will confirm the validity of the
law
both theoretically and experimentally.

\section{Model Study with Catalytic Reaction Network}

In order to search for universal laws of replicating cells which are
independent of details,
we employ a simple reaction network model following Ref.8.
Consider a cell consisting of a variety of chemicals.
The internal state of the  cell can be represented
by a set of numbers $(n_1,n_2,\cdots ,n_k)$, where
$n_i$ is the number of molecules of the chemical species $i$ with $i$
ranging from $i=1$ to $k$.  We choose a randomly generated catalytic
network among these $k$ chemicals, where each reaction from some
chemical $i$ to some chemical $j$ is assumed to be catalyzed by a
third chemical $\ell$, that is  $(i + \ell \rightarrow j + \ell)$.
For simplicity all the reaction coefficients were chosen to be equal
while
the connection paths of this catalytic network were chosen randomly such
that
the probability of any two chemicals $i$ and $j$ to be connected is
given by the connection
rate $\rho$.

Some chemicals diffuse between the cell and the environment
with the diffusion coefficient $D$.
Among the penetrable chemicals, nutrients without catalytic activity are supplied
by the environment.
Through the catalytic reactions to synthesize impenetrable
chemicals, the total amount of chemicals $N= \sum_i n_i$ in a cell can
increase, and
when the total amount of chemicals is beyond some threshold $N_{max}$,
the cell is divided into two.  This growth and division processes
is repeated.
In the numerical simulations, we randomly pick up a pair of molecules in
a cell, and
transform them according to the reaction network.
In the same way, diffusion through the membrane
is also computed by randomly choosing molecules inside the cell and
nutrients in the environment
(see Ref.8 for details of the model).

As the diffusion coefficient $D$
is increased, the growth speed of the cell increases up to a
critical value $D_c$ after which
the cell ceases growing because
the flow of nutrients from the environment is so fast that
the internal reactions transforming them into chemicals sustaining its
`metabolism' cannot keep up.
As discussed in Ref.8, the intra-cellular dynamics
at $D \sim D_c$ is biologically relevant due to
the efficient recursive growth for this value and
its statistical property of chemical abundances, i.e.,
the power-law distribution with exponent -1 which is confirmed
experimentally for almost all cells we investigated.
Accordingly, this simple model captures enough basic properties of a
cell
to adequately reflect universal statistical
properties and we therefore choose it to measure the distribution of
each chemical's abundance over many cells, by sampling them over a large
number of divisions.
~\\

\noindent
{\bf Results of Simulations.}\\
In Fig.1, the number distributions of several chemicals for
$D \sim D_c$ are plotted \cite{continuous_limit}.
Here we measure the number of molecules of each chemical
when a cell divides into two and
the distribution indeed is nearly log-normal. I.e.\

\begin{equation}
P(n_i) \approx exp(-\frac{(logn_i-log\overline{n_i})^2}{2\sigma}),
\end{equation}

\noindent
where $\overline{n_i}$ indicates the average of $n_i$ over the cells.

This log-normal distribution holds for the abundances of all chemicals
except for those that are supplied externally as nutrients which obey
the standard Gaussian
distribution.  In other words, those molecules that are reproduced in a
cell
obey a log-normal distribution, while nutrients that are just transported
from the outside of a cell follow a Gaussian distribution.

Why would the log-normal distribution law generally hold, in spite of
the fact that the central limit theorem implies that the addition of
several fluctuation terms should
lead to a more Gaussian distribution? 
This can be understood by considering that the recursive production
process is a cascade reaction 
near the critical state $D \approx D_c$ \cite{Zipf}.  At this point,
a small part of the possible reaction pathways is dominant and organized
in
a cascade of catalytic reactions so that
a chemical in the $i$-th group  is catalyzed by a chemical in the
$(i+1)-th$ group.
In other words, a "modular structure" with groups of  successive
catalytic reactions
is self-organized in the network.
In this cascade of catalytic reactions,
 fluctuations propagate ``multiplicatively''; for example,
the concentration fluctuation of a
chemical in the $(i+2)$-th group influences multiplicatively that of the
$(i+1)$-th group,
which then influences multiplicatively
that of the $i$-th group, and so forth.
The result of this multiplicative effect is the log-normal distribution
of $n_m$ \cite{auto_catalytic}.  
Note that, at the critical state with which we are concerned,
this cascade of catalytic reactions continues over all chemical species
that are reproduced,
and that the log-normal distribution holds clearly.
The importance of cascade process for log-normal distribution is
also studied in the theory of turbulence \cite{turbulence}, where energy cascade leads to
multiplicative creation of vortices.  In the present case,
cascade in the catalytic reaction is essential to the log-normal distribution.

The width of the distribution for each chemical shown in Fig.1 looks
almost
independent of its average.
This suggests a connection between the fluctuations and the averages of
the chemicals.
We therefore plotted the standard deviation of each chemical
$\sqrt{\overline{(n_i-\overline{n_i})^2}}$ as a function of the average
$\overline{n_i}$
in Fig.2 and indeed found a linear relationship between the standard
deviation ({\em not } the variance)
and the average number of molecules.
This can be understood by considering the
steady growth and cascade structure of the catalytic reactions.
Take two chemicals $i$ and $j$, one of which ($j$)
catalyzes the synthesis of the other in the cascade.
During the steady growth phase of a cell,
the synthesis and conversion of chemical $i$ should be balanced, i.e.,
$n_j \times A -n_i \times B =0$,  where $A$ and $B$ are average
concentrations of
other chemicals involved in the catalytic reaction.
The average concentration then satisfies $\overline{n_j} /\overline{n_i}
= A / B $.
Taking into account that the relation remains satisfied as $n_i,n_j...$
increase
while the cell grows, it is rather natural to assume that the
relationship holds for the fluctuations of the average as well:
$<\delta n_j>^2 /<\delta n_i>^2 =(A/B)^2=\overline{n_j}^2
/\overline{n_i}^2$.
Hence  the variance is expected to be
proportional to the square of the mean, yielding the
linear relationship between the mean and the standard deviation.

A linear relationship is also found as with regards to the variation of
the chemical abundances when changing the external conditions.  For
example,
we  computed the change from  $\overline{n_i}$ to $\overline{n'_i}$ by
varying the concentrations of the supplied nutrients.
The variation $|\overline{n'_i} -\overline{n_i}|$
is again found to be proportional to $\overline{n_i}$ for each chemical
$i$, similar
to the data plotted in Fig.2.

Through extensive simulations of a variety of related models, we have
confirmed that
the discovered laws
hold generally and do not rely on the details of the model,
such as the kinetic rules of the reactions, or 
the structure of reaction network including networks with heterogeneous 
path connectivity as scale-free topology.
They are universal properties of replicating cellular systems
near the critical state $D \approx D_c$.

Of course, the arguments for the two laws thus far are based on the
recursive production of
a cell.  In general, there can be deviations
from the two laws, if the steady growth condition for a cell is not
satisfied.
In the present case, for example, recursive production is not possible
when
the parameter $D$ is much smaller than $D_c$ as  all the possible
reaction pathways
occur with similar weights and the cascade of catalytic reactions is
replaced by a random reaction network.
In this case, the fluctuations of the molecule numbers are highly
suppressed, and the
distributions are close to normal Gaussian.
The multiplicative stochastic process supported by the cascade of catalytic reaction
is replaced by several parallel catalytic processes, and 
the central limit theorem for the addition of stochastic processes would
lead to Gaussian distribution of chemicals.

Furthermore, we have confirmed numerically that
the variance
(not the standard deviation) increases linearly with the average
concentrations.
In other words, the "normal" behavior expected from the central limit
theorem is observed.

\section{Experiment}

\noindent
Now, we report experimental confirmations of the
two basic laws on the distributions of abundances.
Recalling that the laws are expected to hold for the abundances of a protein synthesized
within cells with recursive (steady) growth, we measured the
distribution of the protein abundances in {\sl Escherichia coli} that are
in the exponential phase of growth, i.e., in a stage of steady growth \cite{network}.
To obtain the distribution of the protein abundances, 
we introduced the fluorescent proteins with appropriate promoters into the cells, 
and measured the fluorescence intensity by flow cytometry.
To demonstrate the universality of the laws, 
we have carried out several sets of experiments by using a variety of promoters and also
by changing places that the reporter genes are introduced (i.e., on the plasmid and on the genome).

The detailed experimental procedures are as follows.
~\\

\noindent
{\bf Methods of Experiments.}

{\bf Plasmids and Strains.}  
Reporter plasmids were constructed by subcloning
tetA promoter from pASK-IBA3 (Sigma Genosys) and {\sl egfp} gene (BD
Biosciences Clontech) into pPROTet.E 6xHN(BD Biosciences Clontech) and
{\sl dsred.t4} gene coding red fluorescence protein (RFP)\cite{DsRed} 
into pTrc99A (Amersham Biosciences). {\sl E. coli} strain
OSU2, a derivative of DH5$\alpha$ that lacks glutamine synthetase gene, was
transformed with these reporter plasmids. {\sl E. coli} strain OSU5 was
constructed by replacing {\sl glnA} gene with {\sl tetA} promoter and {\sl gls-h} \cite{A3}
fused with {\sl gfpuv5} \cite{A2} gene by homologous recombination \cite{A1}.

{\bf Culture and Measurements.}
Cultures of strain OSU2 with reporter plasmid
were grown in LB medium with 100 $\mu$g/ml ampicilline for 6h at 37\degC. To
obtain the high expression level of RFP, the culture was grown to
mid-exponential growth and then induced with 1 mM isopropyl-¦Â-
D-thiogalactoside (IPTG) for 3h at 37\degC. {\sl E. coli} OSU5 was grown in
minimal medium (0.1 M Sodium L-Glutamate Monohydrate, 4g/l glucose,
10.5g/l $\mathrm{K_{2}HPO_{4}}$, 4.5g/l $\mathrm{KH_{2}PO_{4}}$, 50mg/l $\mathrm{MgSO_{4}7H_{2}O}$, 50mg/l thiamine HCl,
10$\mu$M $\mathrm{FeSO_{4}7H_{2}O}$, 0.5$\mu$M $\mathrm{CaCl_{2}}$, micronutrient solution \cite{A4}, 25$\mu$g/ml 
kanamycine ) for 24 h at 37\degC. 
All expression data were collected using COULTER\textregistered
EPICS\textregistered ELITE flow cytometer with a 488-nm argon excitation
laser and bandpass filter at 525 $\pm$ 25 nm for GFP fluorescence and
600-nm dichroic filter for RFP fluorescence. For each culture, 10,000
events were collected. 
We have confirmed that it is within its dynamics range, 
by using commercialized beads with known-amount of fluorescent dyes.
All flow data were converted to text format using
WinMDI Version 2.8.
~\\

\noindent
{\bf Results of Experiments.}

In Fig.3, we have plotted the distributions of the emitted fluorescence
intensity
from {\sl Escherichia coli} cells with the reporter plasmids containing
either
EGFP (enhanced green fluorescent protein) under the control of the tetA
promoter without repression, or DsRed.T4 (monomeric red fluorescent protein) 
under the control of the trc promoter
with and without IPTG induction \cite{IPTG}.
In general, the fluorescence intensity (the abundance of the protein) 
increases with the cell size.
To avoid the effect of variation of cell size, which may also obey log-normal distribution, 
we normalized the fluorescence intensity by the volume of each cell. 
Here we adopt the forward-scatter (FS) signal from the flow cytometry,
to estimate the cell volume.
Indeed, by plotting data of the fluorescence intensity versus 
FS signal, the two are proportional on the average.
(The data points are distributed around the proportionality line between 
the two, as are generally observed for the plot
of fluorescence intensity by flow cytometry).
Hence, we normalized
the fluorescence intensity by dividing the FS signal.
Fig.3 is the distribution of this normalized fluorescence intensity.
Note that all these data are fitted well by log-normal, rather than
Gaussian, distributions,
even though each of the expressions is controlled by a different
condition of the promoter.

To clarify the existence of larger tail for greater abundances of proteins, 
in Fig. 4, we have plotted the distribution of protein abundances 
both in the logarithmic scale and in the normal scale.
The abundances of fluorescent protein expressed from the chromosome is
also found to obey the log-normal distribution, as shown in Fig.4.
Here, the data are obtained by {\sl Escherichia coli} cells with an expression of
glutamine
synthetase (GS) fused to GFP in the chromosome, whose expression is
controlled by
the upstream tetA promoter.
In Fig.4, plotted is the distribution of fluorescence intensity again normalized by the cell volume.
As can be seen, when using the logarithmic scale (a),
the distribution is roughly symmetric and close to Gaussian, while when
using
the normal scale (b), the distribution has a larger
tail on the side of greater abundances. 
The fact that the log-normal distributions are also
observed when genes are located on the genome indicates that 
the nature of log-normal distribution is not due to the variation in plasmid copy number.
We have also examined
several other cases using different reporter genes both
on the plasmids and on the genome,
and obtained similar results supporting the universality of
the log-normal distributions.
It is furthermore interesting to note that the abundances of the
fluorescent proteins, reported in the literature so far,
are often plotted with a logarithmic scale\cite{Collins}.

It should be noted that the log-normal distribution of protein abundances
is observed when the {\sl E. coli} are in the exponential phase of growth, i.e.,
when the bacteria are in the steady growth stage.  For other phases of growth without
steady growth, the distribution is found to be often deviated from the
log-normal distribution, and sometimes show distribution with double peaks,
as will be reported in future.   Note that the theory also supports the
log-normal distribution for the steady growth case only.

As for the linear relationship between the variation and the average,
Banerjee et al. \cite{Banerjee} recently reported that 
the standard deviation of gene expressions in cell population is proportional to the average expression level,
which support the relationship discovered in our study.
However, using a different system, Ozbudak et al. \cite{Ozbudak} showed that 
the standard deviation of gene expression is not proportional to its average, but 
the variance increases linearly with the average.
However, in these studies, the dynamic ranges of the measurements are relatively narrow
and the growth conditions of cells are not kept precisely in the log phase,  so that the
condition for the steady growth is not satisfied.
To confirm the linear relationship between the standard deviation and the average, 
further experimental studies with a wide dynamic range of measurements and 
precise control of the steady cellular growth are required.

\section{Discussion}

To sum up, we have reported universal laws on the distributions of
chemical abundances in cells with steady growth.
First, the distribution of chemical abundance obeys the log-normal distribution
due to the multiplicative propagation of fluctuations.
Second, there is a linear relationship between 
the average and standard deviation of chemical abundances.
Since the laws generally appear near the critical state $D \approx D_c$, 
and the dynamics at the critical state provides 
a faithful and efficient self-reproduction of a cell \cite{Zipf}, 
it is natural to conclude that cells generally hold these laws.
We have also shown the experimental confirmations of the law using {\sl E. coli} cells, 
where the number distributions of fluorescent proteins obey the log-normal distribution 
independent of the conditions of promoters and locations of reporter genes.
The ubiquity of the discovered
log-normal distributions can be
a solid basis for the study of fluctuations in cells.
It should be stressed that the log-normal distribution of chemical
abundances implies that the average magnitude of the fluctuations is
much larger than what one would observe for the normal distribution.
However, at the present time, analysis of cellular heterogeneity
mostly relies on the Gaussian distribution in the abundance of chemicals.
Hence, our discovery for the ubiquity of log-normal distribution sets drastic
and essential changes to future studies concerning with fluctuations
in cellular dynamics. 

Clearly, these
laws bear relevance to adaptation and evolution\cite{Sato} as well since the
role of phenotypic fluctuations cannot be neglected.
With these two laws in mind, it is therefore
important to further study how cells maintain their functions
and replicate themselves successfully
despite being subjected to such large fluctuations, and to search for
possible relationships between the topology of the reaction
networks and the fluctuations in intra-cellular reaction dynamics.
In relation,
the search for some `exceptional' chemicals that do not follow
the log-normal distribution that may be located at specific positions in
the network will be interesting. Indeed, the log-normal distribution appears as a result of
multiplicative propagation of noise in cascade catalytic reaction process, and
the distribution could be sharpened by interference of parallel reaction processes, 
including negative feed-back loop.

Finally, we note again that the observed laws hold for a cell that grows recursively.
For cells undergoing the change of the states, the distribution
could be distorted or have double peaks, as observed when the condition of culture
is changed or in the course of cell differentiation.  The present two 
laws could be a basis for studying such change of distribution as a
measure of biological plasticity. 

We would like to thank K. Sato and K. Ohnuma 
for stimulating discussions and 
Frederick H. Willeboordse for critical reading of the manuscript.
Grant-in-Aids for Scientific Research from
the Ministry of Education, Science and Culture of Japan
(11CE2006).

\newpage

\noindent
{\bf \large Figure regends}
~\\

\noindent {\bf Fig.1.} 
The number distribution of the molecules of chemical
abundances of our model.
Distributions are plotted
for several chemical species with different average molecule numbers.
The data were obtained by observing 178800 cell divisions.
~\\
~\\
\noindent {\bf Fig.2.} 
Standard deviation versus average number of molecules.
Using the same data set and parameters as for Fig.1,
the relationship between
the average and standard deviation is plotted for all chemical species.
The solid line is for reference.
~\\
~\\
\noindent {\bf Fig.3.} 
The number distribution of the proteins measured by fluorescence
intensity, normalized by the cell volume.
Distributions are obtained from three {\sl Escherichia coli} cell
populations
containing different reporter plasmids (see text).
Note that, although the IPTG induction changes the average fluorescence
intensity,
 both the distributions (with and without the induction) can be fitted
by log-normal distributions well.
~\\
~\\
\noindent {\bf Fig.4.} 
The distribution of the fluorescence intensity normalized by the cell
volume, plotted (a) with a logarithmic scale and (b) with a normal scale.
Data are obtained from a population of isogenic bacterial
cells with an expression of GFP-GS fusion protein in the chromosome.
It is clear that the distribution with the logarithmic scale is
symmetric and close to a Gaussian form.

\newpage

\begin{figure}[tbp]
\begin{center}
\includegraphics[width=12cm,height=9cm]{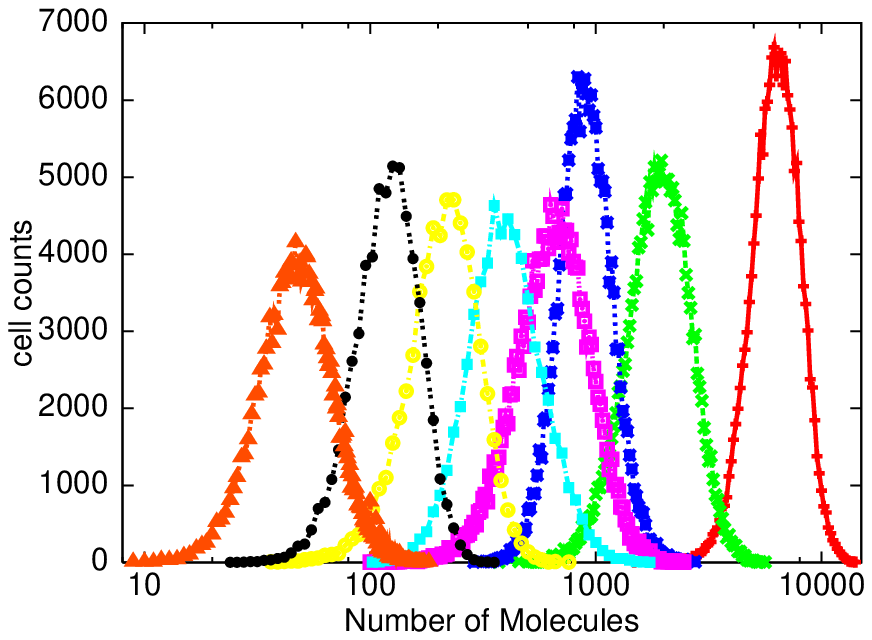}
\end{center}
\caption{
}
\end{figure}

\begin{figure}[tbp]
\begin{center}
\includegraphics[width=12cm,height=9cm]{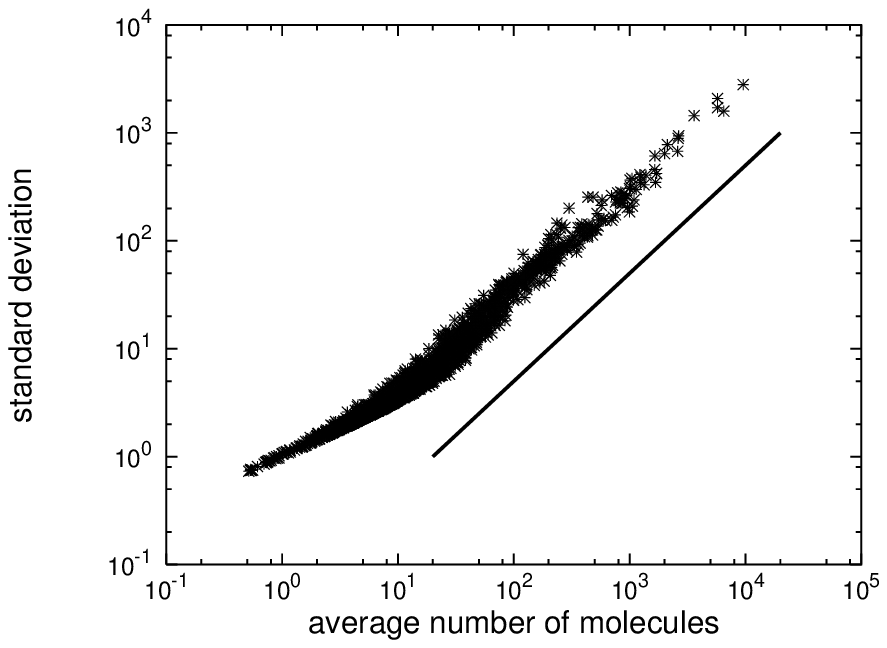}
\end{center}
\caption{
}
\end{figure}

\begin{figure}[tbp]
\begin{center}
\includegraphics[width=12cm,height=9cm]{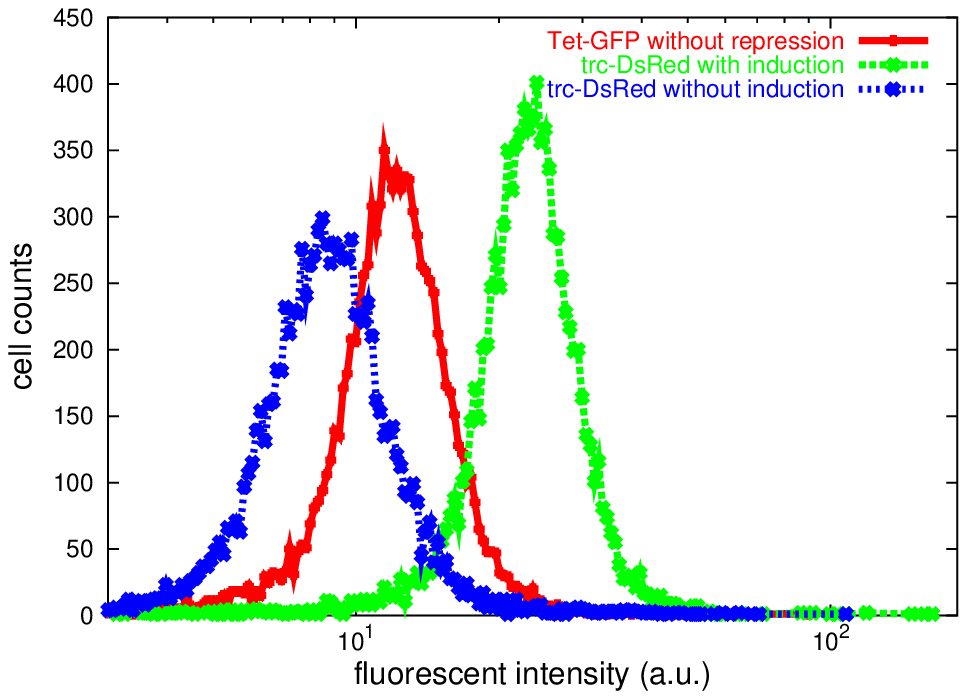}
\end{center}
\caption{
}
\end{figure}

\begin{figure}[tbp]
\begin{center}
\includegraphics[width=10cm,height=14cm]{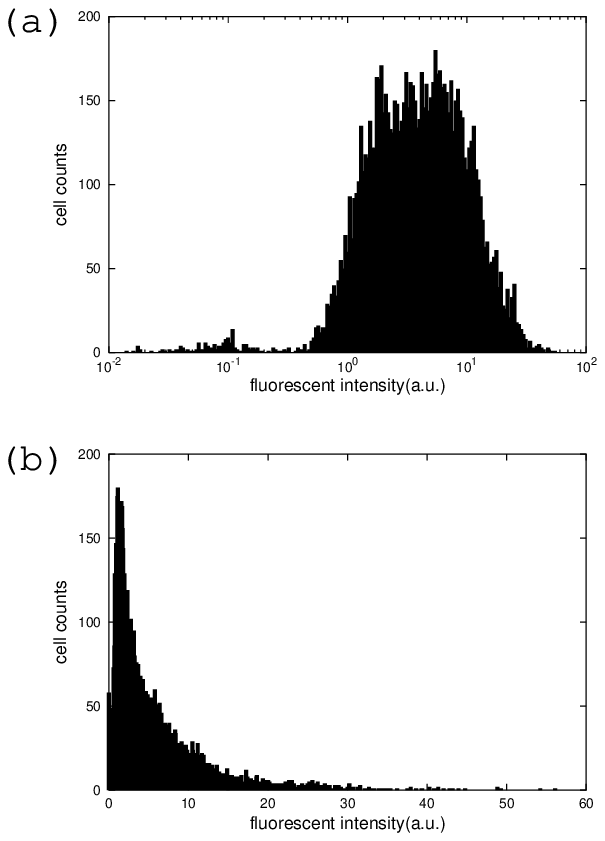}
\end{center}
\caption{
}
\end{figure}


\begin{thebibliography}{999}
\bibitem{continuous_limit}
In the simulation used to generate Fig.1,
the total number
of molecules $N$ is much larger than the number of chemical species $k$.
In this case, the population ratios $\{ n_i/N \}$ generally fall
into a unique steady state and fluctuate around it, while the daughter
cells
inherit the chemical compositions of their mother cells.
On the other hand, when $k>N_{max}$,
the population ratios do not
settle down and can change from generation to generation.
In this case, the number distribution of each chemical does not obey the
log-normal
distribution as shown in Fig.1, but seems to follow a Poisson-like
distribution.
\bibitem{auto_catalytic}
The heuristic argument on the log-normal distribution in \S 2
intended for  auto-catalytic processes is not
directly applicable here, since in the present study, catalysts are generally other molecule
species, and such direct auto-catalytic process is not particularly
enhanced.  Still,  some revision might be possible,
since a set of chemical species could work as an auto-catalytic set
to synthesize the same set of chemicals, for a recursive production of a cell.
\bibitem{network}
Concentration dynamics of a given protein is controlled by many factors, 
including concentrations of transcriptional factors, several subunits of
RNA polymerases, ribosomal proteins, RNA and protein degradations, 
number of plasmids, and so forth.
Since these factors regulate each other through chemical reactions, 
it is natural to describe the dynamics of a protein concentration 
using a network of reactions as our model.
\end{thebibliography}
\end{document}